\documentclass{elsarticle}
\usepackage{amsfonts,amsmath,amssymb}
\usepackage{float}
\usepackage{graphicx}
\usepackage[labelfont=bf]{caption}
\usepackage{booktabs}
\usepackage{url}

\usepackage{lineno}
\modulolinenumbers[5]

\journal{MATCOM (Elsevier)}

\biboptions{sort&compress}

\begin{document}

\begin{frontmatter}

\title{Stochastic kinetics of the circular gene hypothesis: feedback effects and protein fluctuations}

\author[ccss]{R.R. Wadhwa}

\author[wigner]{L. Zal\'anyi}

\author[aoss]{J. Szente}

\author[wigner]{L. N\'egyessy}

\author[ccss,wigner]{P. \'Erdi\corref{cor}}
\ead{perdi@kzoo.edu}
\cortext[cor]{Corresponding author: + 1 (269) 337-5720}

\address[ccss]{Center for Complex Systems Studies, Kalamazoo College, 1200 Academy Street, Kalamazoo, MI 49006, USA}
\address[wigner]{Wigner Research Centre for Physics, Hungarian Academy of Sciences, Budapest, Hungary}
\address[aoss]{Atmospheric, Oceanic and Space Sciences, College of Engineering, University of Michigan, Ann Arbor, MI, USA}

%
%
%

\begin{abstract}

Stochastic kinetic models of genetic expression are able to describe protein fluctuations. A comparative study of the canonical and a feedback model is given here by using stochastic simulation methods. The feedback model is skeleton model implementation of the circular gene hypothesis, which suggests the interaction between the synthesis and degradation of mRNA. Qualitative and quantitative changes in the shape and in the numerical characteristics of the stationary distributions suggest that more combined experimental and theoretical studies should be done to uncover the details of the kinetic mechanisms of gene expressions.

\end{abstract}


\end{frontmatter}


\section{Introduction}
\label{sec:intro}

Protein availability is a \emph{condicio sine qua non} of cellular processes and survival, and is determined by gene regulation. Gene regulation contains many biochemical and biophysical processes. While traditional biochemistry adopted a rather rigid deterministic scenario considering the execution of instructions encoded in DNA, chemical reactions taking place at the single cell level are now admittedly better described by stochastic models than by deterministic ones. Reactions in gene expression, such as promoter activity and inactivity, transcription, translation, and decaying of mRNA and proteins are the most important chemical steps. Measurements on stochastic gene expression in single cells with single molecule sensitivity \cite{cai06,elowitz02} implied the necessity of stochastic description \cite{qian13}. Since our goal here is to contribute to the understanding of the nature of protein fluctuations, only stochastic kinetic modeling technique can be relevant. 


The perspective that models of gene expression should have stochastic elements goes back to the pioneering works of D. Rigney and O. Berg \cite{berg78,rigney79b,rigney79a,rigney77}, but these works came too early for mainstream molecular biologists. Stochastic chemical kinetics became the \emph{lingua franca} of modeling gene regulatory networks and related fields twenty years later due to  highly cited papers \cite{arkin98,arkin97}.

Stochastic models proved to be very efficient to study the kinetic mechanisms of genetics and, more generally, systems biological processes \cite{el14}. Measured fluctuations in reactions two sources \cite{elowitz02} (i) \textbf{intrinsic} noise is related to variations in protein levels even in a population of cells with identical genotype and concentrations and states of cellular components, (ii) \textbf{extrinsic} noise due to fluctuations in the amount or activity of molecules involved in the expression of a gene, like RNA polymerase or ribosomes. The reaction system, what might be called the canonical model of gene expression \cite{paulsson05}, belongs to the category of compartmental models. Such systems are characterized by the fact that the activity of one molecular entity is independent of the other entities. In other words, no interaction between any two such entities occurs. Such models can fully be solved (i.e. the time-dependent moments can be calculated) by using the generating function method. More specifically, the different sources of protein fluctuations have been calculated \cite{paulsson05}. Under not too restrictive conditions it was found \cite{cai06,friedman06} that the stationary distribution of the protein fluctuation can be well approximated by gamma distribution. However, as gamma distribution is a very general one, alterations of the reactions system and/or the rate constants may imply changes in the shape and parameters of the stationary distributions. 

Realistic models should take into account feedback, burst, delay, etc. mechanisms too, and some exact results are available for specific families of models \cite{friedman06,bokes12,shah08,kumar14,pendar13}.These models contain bimolecular reaction steps too, so the compartmental kinetic framework based on independent activities cannot be assumed anymore. Linear noise approximation is often used to calculate protein fluctuations, but its reliability for systems containing bimolecular reaction steps is restricted \cite{thomas13}.

Simulation methods (for a recent short review see $2.6$ of \cite{el14}) are appropriate tools to obtain information about the size and nature of fluctuations as they help overcome the limitations of the methods described above.

A recent  conceptually new hypothesis \cite{haimovich13} suggested that gene expression might be circular, since the degradation and synthesis of mRNA seem to be interconnected by a feedback mechanism. Due to the lack of available kinetic data the hypothesis cannot be falsified for the time being. However, as the metabolism of mRNA is better described by some bimolecular reactions, it might affect protein fluctuations. Specifically, by setting a secondary mechanism to promote mRNA synthesis may increase the lifetime ratio of the lifetimes of mRNA and of proteins, which increases protein fluctuation. 
Our question was whether or not the feedback mechanism has a significant effect on protein fluctuations. If yes, it is worth studying the details. 

\section{Biological background}
\label{sec:biobg}

Gene expression is the complicated process of converting genetic information from a DNA sequence into proteins. In eukaryotes, DNA is located in the cell nucleus. Prokaryotes do not have a nucleus, and DNA can be found in the cytoplasm. In prokaryotes there are two main processes in gene expression: \textbf{transcription}\index{transcription} and \textbf{translation}\index{translation}. In eukaryotes, there is an additional process: \textbf{splicing}\index{splicing}.

\textbf{Transcription} is a series of events that use DNA to synthesize messenger RNA (mRNA) by using the enzyme RNA polymerase as a catalyst. The series of events contain (in prokaryotes) \textit{binding}\index{transcription!binding}, \textit{initiation}\index{transcription!initiation}, \textit{RNA synthesis}\index{transcription!RNA synthesis}, \textit{elongation}\index{transcription!elongation}, and \textit{termination}\index{transcription!termination}. Specifically, a \textbf{promoter} is a region of DNA where binding of transcription factor proteins initiate transcription. Eukaryotic transcription is much more complicated, but depends on these basic steps.

\textbf{Splicing} is a modification of the nascent mRNA transcript in which certain nucleotide sequences (\textit{introns})\index{intron} are removed while other sequences (\textit{exons})\index{exon} remain.

\textbf{Translation} is a process in which the \index{genetic code} is read out from the mRNA by the ribosome complex and translated into the amino acid sequence in proteins (with the help of tRNA). It contains more elementary steps, such as \textit{initiation}\index{translation!initiation}, \textit{elongation}\index{translation!elongation}, \textit{translocation}\index{translation!translocation}, and \textit{termination}\index{translation!termination}.

\textbf{Degradation:} although DNA is stable, RNA and protein molecules are subject to degradation. It is an important step in the regulation of gene expression and fluctuation of protein concentration.

A recent hypothesis \cite{haimovich13} suggested that eukaryotic gene expression can be viewed as a circular process, where transcription and mRNA degradation are interconnected. The big question is, ``How could mRNA synthesis in the nucleus and mRNA decay in the cytoplasm be mechanistically linked?'' \cite{braun-young14}. Possible mechanisms of coupling mRNA synthesis and decay  have been analyzed \cite{braun-young14,medina14}. The $5'$ to $3'$  exoribonuclease $xrn1$, a large protein involved in cytoplasmatic mRNA degradations might be a critical component \cite{medina14}, and it may play a dual role in some subprocesses of transcription, namely in initiation and elongation.

Based on these observations about the dual role of $xrn1$ in transcription~\cite{haimovich13,medina14}, a minimal model that takes into account feedback effects has been set by including three more steps: promoter assignment, promoter reassignment, and $xrn1$ dependent transcription.


In this paper, the nature of protein fluctuation in the canonical model and a simple feedback model implementing the dual role of $xrn1$ is studied using stochastic simulations. Based on the results, we predict that the feedback process has significant effects on the fluctuations by the additive effects of the enhanced mRNA fluctuations, so the detailed mechanisms should be studied by combined experimental and modeling studies.


\section{The model}
\label{sec:model}

\subsection{Canonical model}

Gene expression can be modeled as a three-stage process: gene activation, transcription, and translation. These are coupled by the opposite processes of gene inactivation, mRNA degradation, and proteolysis, respectively. Gene expression can be modeled as a reaction system of three chemical species, slightly modified from the existing schematic for the canonical model \cite{bokes12}.
\begin{center}
\begin{tabular} {r r c l}
Gene activation: & inactive gene & $\overset{\lambda_1}{\xrightarrow{\hspace*{0.7cm}}}$ & active gene \\
Gene inactivation: & active gene & $\overset{\lambda_2}{\xrightarrow{\hspace*{0.7cm}}}$ & inactive gene \\
Transcription: & active gene & $\overset{\rho_1}{\xrightarrow{\hspace*{0.7cm}}}$ & active gene + mRNA \\
mRNA degradation: & mRNA & $\overset{\rho_2}{\xrightarrow{\hspace*{0.7cm}}}$ & $\varnothing$ \\
Translation: & mRNA & $\overset{\gamma_1}{\xrightarrow{\hspace*{0.7cm}}}$ & mRNA + protein \\
Proteolysis: & protein & $\overset{\gamma_2}{\xrightarrow{\hspace*{0.7cm}}}$ & $\varnothing$
\end{tabular}
\end{center}
In this system, all the reactions are first order. This reaction system can be alternatively defined using the number of each chemical species present in a cell. Here $n_1$, $n_2$, $n_3$, and $n_4$ represent the number of inactive genes, active genes, mRNAs, and proteins in the cell, respectively \cite{paulsson05}.
\begin{center}
\begin{tabular} {r r c l}
Gene activation: & ($n_1,n_2$) & $\overset{\lambda_1 n_1}{\xrightarrow{\hspace*{1cm}}}$ & ($n_1-1,n_2+1$) \\
Gene inactivation: & ($n_1,n_2$) & $\overset{\lambda_2 n_2}{\xrightarrow{\hspace*{1cm}}}$ & ($n_1+1,n_2-1$) \\
Transcription: & $n_3$ & $\overset{\rho_1 n_2}{\xrightarrow{\hspace*{1cm}}}$ & $n_3+1$ \\
mRNA degradation: & $n_3$ & $\overset{\rho_2 n_3}{\xrightarrow{\hspace*{1cm}}}$ & $n_3-1$ \\
Translation: & $n_4$ & $\overset{\gamma_1 n_3}{\xrightarrow{\hspace*{1cm}}}$ & $n_4+1$ \\
Proteolysis: & $n_4$ & $\overset{\gamma_2 n_4}{\xrightarrow{\hspace*{1cm}}}$ & $n_4-1$
\end{tabular}
\end{center}
To determine the value of the rate constants, we refer to experimental results, using $\emph{E. coli}$ as our model organism. The half-life of mRNA in $\emph{E. coli}$, calculated as the natural logarithm of 2 divided by the rate constant for mRNA degradation, is between 3 min and 8 min \cite{bokes12}. Using a timescale measured in seconds, we choose the average half-life of an mRNA molecule in the simulation to be 300 s. This leads to a value of $\text{ln(2)}/300 \approx 0.00231$ for $\rho_2$. 

There are indications from experimental data that $\rho_1/\rho_2$ is variable in $\emph{E. coli}$, ranging from 1 to a few dozen - we assume a value of 10 for this ratio, which implies $\rho_1 \approx 0.0231$ \cite{bokes12,lu07}. It has also been experimentally determined that for proteins in $\emph{E. coli}$, the average value of $\gamma_1/\gamma_2$ is 540 \cite{bokes12,lu07}. For this simulation, we choose $\gamma_1=0.14$ as it approaches a stationary state with sufficient speed. For the sake of simplicity, we assume there exists only a single copy of the gene we are interested in. We also choose $\lambda_1=1$ and $\lambda_2=7$, although we shall see that the choice of values for $\lambda_1$ and $\lambda_2$ is arbitrary. The ratio $\lambda_1/(\lambda_1+\lambda_2)$ indicates the proportion of time for which a gene is active~\cite{paulsson05}, and is what truly matters.

The initial value of $(n_1,n_2,n_3,n_4)$ for the reaction system was $(1,0,0,0)$.

\subsection{Feedback model: the role of $xrn1$}

Gene expression involving $xrn1$ requires a model with more reactions to be accurately modeled. Using the biological background given in Section~\ref{sec:biobg}, we give the following model to account for feedback due to $xrn1$. Here $n_5$, $n_6$, and $n_7$ represent the number of $xrn1$ molecules, $xrn1$ complexes, and $xrn1$ binding to the promoter in the cell, respectively.
\begin{center}
\begin{tabular} {r r c l}
Gene activation: & $(n_1,n_2)$ & $\overset{\lambda_1 n_1}{\xrightarrow{\hspace*{1cm}}}$ & $(n_1-1,n_2+1)$\\
Gene inactivation: & $(n_1,n_2)$ & $\overset{\lambda_2 n_2}{\xrightarrow{\hspace*{1cm}}}$ & $(n_1+1,n_2-1)$\\
Transcription: & $n_3$ & $\overset{\rho_1 n_2}{\xrightarrow{\hspace*{1cm}}}$ & $n_3+1$\\
Promoter assignment: & $(n_2,n_6,n_7)$ & $\overset{\omega_1 n_2 n_6}{\xrightarrow{\hspace*{1cm}}}$ & $(n_2-1,n_6-1,n_7+1)$\\
Promoter reassignment: & $(n_2,n_5,n_7)$ & $\overset{\omega_2 n_7}{\xrightarrow{\hspace*{1cm}}}$ & $(n_2+1,n_5+1,n_7-1)$\\
$xrn1$ dep. transcription: & $n_3$ & $\overset{\rho_3 n_7}{\xrightarrow{\hspace*{1cm}}}$ & $n_3+1$\\
Translation: & $n_4$ & $\overset{\gamma_1 n_3}{\xrightarrow{\hspace*{1cm}}}$ & $n_4+1$\\
mRNA degradation: & $(n_3,n_5,n_6)$ & $\overset{\rho_2 n_3 n_5}{\xrightarrow{\hspace*{1cm}}}$ & $(n_3-1,n_5-1,n_6+1)$\\
Proteolysis: & $n_4$ & $\overset{\gamma_2 n_4}{\xrightarrow{\hspace*{1cm}}}$ & $n_4-1$\\
\end{tabular}
\end{center}

It is experimentally supported that the rate constant of $xrn1$ dependent transcription is equal to the rate constant of transcription, implying that $\rho_3=\rho_1=0.0231$ \cite{haimovich13}. The effects of varying values of $\omega_1$ and $\omega_2$ on the resulting protein distribution are investigated in Section~\ref{sec:results}. The remaining rate constants in the feedback model have values identical to the corresponding rate constants in the canonical model.

The initial value of $(n_1,n_2,n_3,n_4,n_5,n_6,n_7)$ for the reaction system was $(1,0,0,0,10,0,0)$.

\begin{center}
\begin{minipage}{0.95\textwidth}
	\begin{minipage}[c]{0.59\textwidth}
		\centering
		\includegraphics[width=\textwidth]{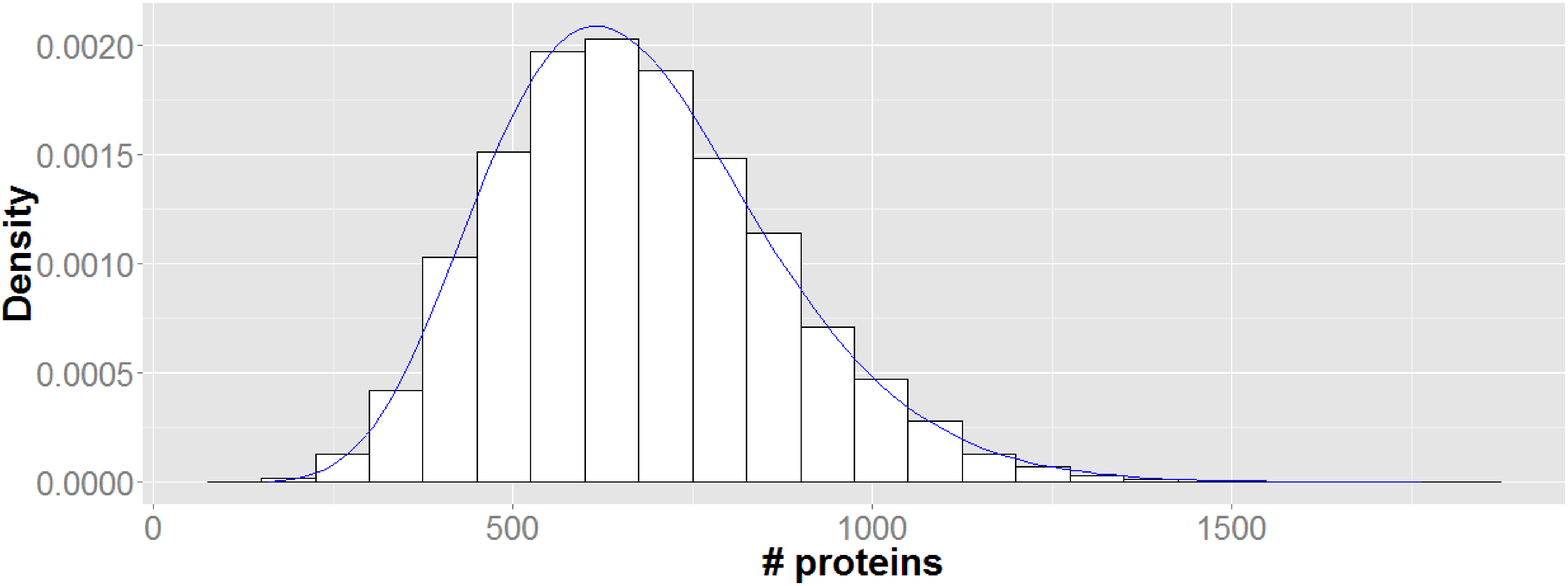}
		\captionof{figure}{\footnotesize{The distribution of proteins in the canonical model at $t=100,000$ s with $n=10,000$ stochastic trajectories. The plotted line is the best-fit gamma distribution with a shape parameter of $11.59 \pm 0.16$ and a rate parameter of $0.0172 \pm 0.0002$. The expected value is equal to $674.54$ and the standard deviation of the distribution is $196.80$.}}
		\label{fig:orig-can}
	\end{minipage}
	\hfill
	\begin{minipage}[c]{0.39\textwidth}
		\centering
		\begin{tabular}{c c}\hline
			Rate Constant & Value \\
			\hline
			$\lambda_1$ & 1.0 \\
			$\lambda_2$ & 7.0 \\
			$\rho_1$ & 0.0231 \\
			$\rho_2$ & $\rho_1/10$ \\
			$\gamma_1$ & 0.14 \\
			$\gamma_2$ & $\gamma_1/540$ \\
			\hline
		\end{tabular}
		\captionof{table}{\footnotesize{The rate constants used to produce the protein distribution of the canonical model on the left. $\rho_1$ has been experimentally determined, as have the ratios $\rho_1/\rho_2$ and $\gamma_1/\gamma_2$~\cite{bokes12,lu07}.}}
		\label{table:orig-can}
	\end{minipage}
\end{minipage}
\end{center}

\subsection{Simulation method}
All simulations for this paper were conducted in the Cain interface developed by Sean Mauch \cite{mauch11}. Realizations of the stochastic process were produced using Gillespie's direct method \cite{gillespie77}. Considering a reaction system as a set of ordinary differential equations as defined by the laws of mass action kinetics, numerical integration using the Cash-Karp variant of the Runge-Kutta method was conducted to produce deterministic trajectories \cite{cashkarp90}. All histograms and plots were produced using the ggplot2 package in the R programming language with the multiplot function from the Cookbook for R website \cite{cookbook,wickham09}. Function fitdistr( ) in the MASS package from the R programming language was used to find the best fit parameters for the distributions.

\begin{center}
\begin{minipage}{0.95\textwidth}
	\begin{minipage}[c]{0.59\textwidth}
		\centering
		\includegraphics[width=\textwidth]{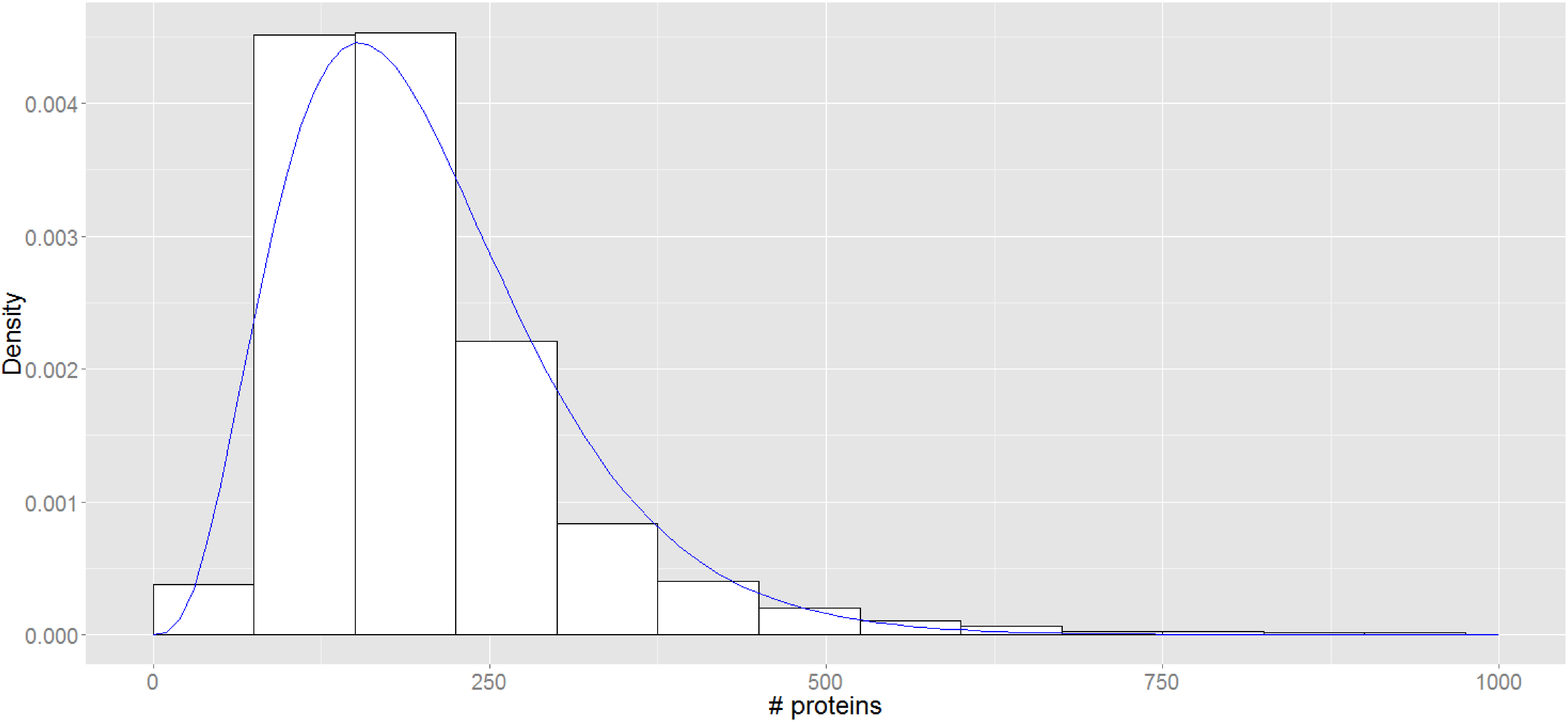}
		\captionof{figure}{\footnotesize{The distribution of proteins in the feedback model at $t=100,000$ s with $n=10,000$ stochastic trajectories. Plotted lines are the best-fit gamma distribution with a shape parameter of $4.089 \pm 0.06$ and a scale parameter of $0.020 \pm 0.0003$. The expected value is equal to $203$ and the standard deviation of the distribution is $124.29$.}}
		\label{fig:orig-fb}
	\end{minipage}
	\hfill
	\begin{minipage}[c]{0.39\textwidth}
		\centering
		\begin{tabular}{c c}\hline
			Rate Constant & Value \\
			\hline
			$\lambda_1$ & 1.0 \\
			$\lambda_2$ & 7.0 \\
			$\rho_1$ & 0.0231 \\
			$\rho_2$ & $\rho_1/10$ \\
			$\rho_3$ & $\rho_1$ \\
			$\omega_1$ & 1 \\
			$\omega_2$ & $3\rho_1/2$ \\
			$\gamma_1$ & 0.14 \\
			$\gamma_2$ & $\gamma_1/540$ \\
			\hline
		\end{tabular}
		\captionof{table}{\footnotesize{The rate constants used to produce the protein distribution of the feedback model on the left. $\rho_1$ and $\rho_3$ have been experimentally determined, as have the ratios $\rho_1/\rho_2$ and $\gamma_1/\gamma_2$~\cite{bokes12,lu07}.}}
		\label{table:orig-fb}
	\end{minipage}
\end{minipage}
\end{center}

\section{Simulation Results}
\label{sec:results}

Stationary protein distributions for both models were plotted as histograms, and fitted with gamma distributions. Figure~\ref{fig:orig-can} shows the steady state protein distribution for the canonical model with the typical parameter values given in Table~\ref{table:orig-can}. Figure~\ref{fig:orig-fb} shows the steady state distribution for the feedback model with the typical parameter values given in Table~\ref{table:orig-fb}. The increased skew in the steady state protein distribution of the feedback model as compared to the canonical model is observed in all the steady state distributions explored.

\subsection{Varying the rate of gene (in)activation: $\lambda_1$ and $\lambda_2$}
\label{ssec:l1l2}

As stated in Paulsson (2005), the kinetic dynamics of gene activation and inactivation can be approximately modeled as a random telegraph process, where each gene independently switches on with rate $\lambda_1$ and switches off with rate $\lambda_2$. The probability of a gene being on is given by $P_\text{on}$ = $\lambda_1/(\lambda_1+\lambda_2)$. Correspondingly, the expected number of proteins in steady state should be proportional to $P_\text{on}$. This pattern is confirmed by Figure~\ref{fig:det-stoch-can}, wherein the expected number of proteins at steady state for the deterministic and stochastic results is linearly proportional to $P_\text{on}$. Due to the presence of the relationship described by Paulsson and confirmation in the canonical model, we arbitrarily choose the values $\lambda_1$=1 and $\lambda_2$=15 when exploring the more complicated feedback model. Results for other values of $\lambda_1$ and $\lambda_2$ can be extrapolated by finding the value of $P_\text{on}$ for the scenario in question.

\begin{table}
\begin{center}
	\begin{tabular}{|r| c c c c}
	\hline
	 & 1/16 & 1/8 & 1/4 & 1/2 \\
	\hline
	0.01 & 337.5 & 675 & 1350 & 2700 \\
	0.05 & 337.5 & 675 & 1350 & 2700 \\
	0.14 & 337.5 & 675 & 1350 & 2700 \\
	0.50 & 337.5 & 675 & 1350 & 2700 \\
	\end{tabular}

	\caption{\footnotesize{Expected number of proteins as given by deterministic solutions of the canonical model. The horizontal labels show varying values of $\lambda_1/(\lambda_1+\lambda_2)$. The vertical labels show varying values of $\gamma_1$. $\rho_1=0.0231$; $\rho_2=\rho_1/10$; $\gamma_2=\gamma_1/540$.}}
	
	\label{table:det-can}
\end{center}
\end{table}

\begin{table}
\begin{center}
	\begin{tabular}{|r| c c c c}
	\hline
	& 1/16 & 1/8 & 1/4 & 1/2 \\
	\hline
	0.01 & 284.3 $\pm$ 41.0 & 568.2 $\pm$ 58.2 & 1136.8 $\pm$ 83.1 & 2275.1 $\pm$ 116.5 \\
	0.05 & 336.8 $\pm$ 84.8 & 675.6 $\pm$ 120.5 & 1350.7 $\pm$ 173.9 & 2702.0 $\pm$ 241.1 \\
	0.14 & 338.2 $\pm$ 139.6 & 676.4 $\pm$ 194.7 & 1349.7 $\pm$ 274.6 & 2701.0 $\pm$ 387.3 \\
	0.50 & 335.1 $\pm$ 223.1 & 675.9 $\pm$ 326.2 & 1342.3 $\pm$ 455.3 & 2705.2 $\pm$ 642.1 \\ 
	\end{tabular}
	
	\caption{\footnotesize{Expected number of proteins $\pm$ standard deviation of the steady state protein distribution as given by stochastic simulations of the canonical model. The horizontal labels show varying values of $\lambda_1/(\lambda_1+\lambda_2)$. The vertical lables show varying values of $\gamma_1$. $\rho_1=0.0231$; $\rho_2=\rho_1/10$; $\gamma_2=\gamma_1/540$.}}

	\label{table:stoch-can}
\end{center}
\end{table}

\subsection{Varying the translation rate: $\gamma_1$}
\label{ssec:g1}

\begin{figure}
\begin{center}
	\includegraphics[width=\textwidth]{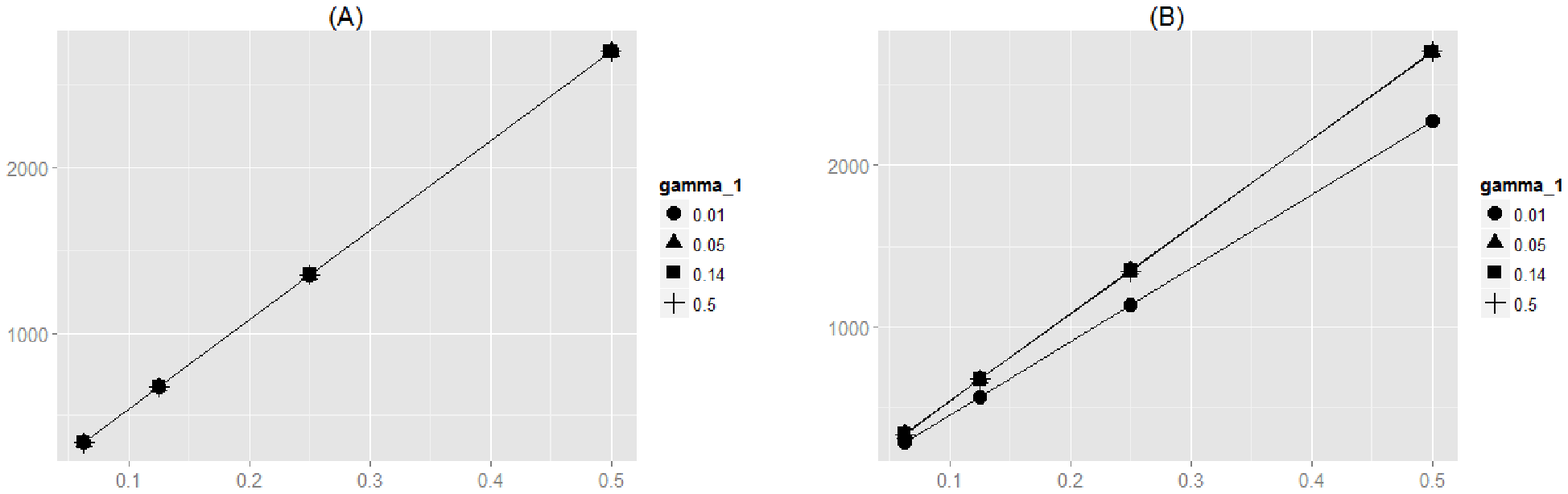}
	\caption{\footnotesize{The x-axis on both plots is $\lambda_1$/($\lambda_1$+$\lambda_2$); the y-axis on both plots is the mean number of proteins at steady state. (A) Deterministic solutions for the canonical model. The plotted series all match exactly (Table~\ref{table:det-can}). (B) Stochastic simulation results for the canonical model. The series for $\gamma_1$ = 0.05, 0.14, and 0.50 closely overlap (Table~\ref{table:stoch-can}).}}
	\label{fig:det-stoch-can}
\end{center}
\end{figure}

Interestingly, the deterministic solutions of the canonical and feedback models are independent of $\gamma_1$ (Figure~\ref{fig:det-stoch-can}A). In the stochastic results of the canonical model, the number of proteins at steady state is lower at $\gamma_1$ = 0.01; for values of $\gamma_1$ = 0.05, 0.14, 0.50 the stochastic results seem to converge to the deterministic solution (Figure~\ref{fig:det-stoch-can}B). However, as seen in Table~\ref{table:stoch-can}, despite this convergence, the standard deviation of the steady state protein distribution increases as $\gamma_1$ increases. These results suggest that using a single value for the number of proteins at steady state as given by the deterministic solution becomes an increasingly worse approximation of the true distribution of proteins in a cell as the translation rate increases.

\begin{figure}
\begin{center}
	\includegraphics[width=\textwidth]{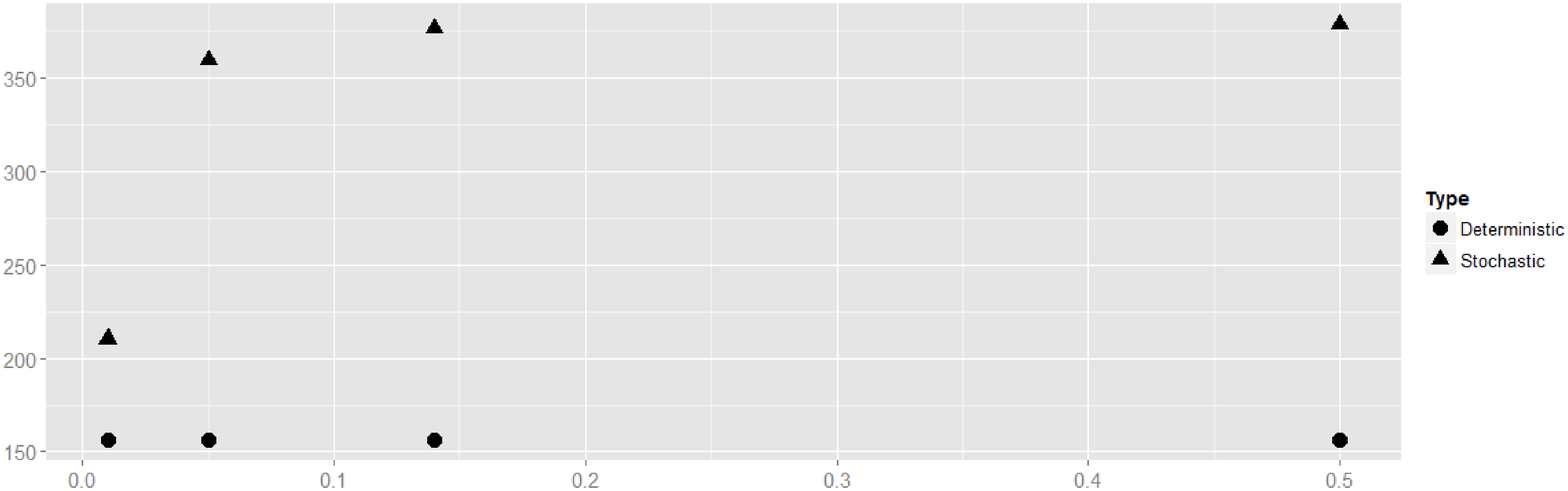}
	\caption{\footnotesize{Deterministic and stochastic solutions for number of proteins at steady state with the given parameter values: $\lambda_1$ =1; $\lambda_2$ = 15; $\omega_1$ = 0.01; $\omega_2$ = 1.25$\rho_1$. The x-axis on both plots is $\gamma_1$; the y-axis on both plots is the mean number of proteins at steady state.}}
	\label{fig:feed-example}
\end{center}
\end{figure}

The convergence pattern does not hold true of the feedback model. Figure~\ref{fig:feed-example} shows an instance where the deterministic and stochastic solutions clearly diverge irrespective of the value of $\gamma_1$. Thus, for the feedback model, the deterministic and stochastic methods give distinct values for the expected number of proteins at steady state.

\subsection{Varying the rate of promoter assignment: $\omega_1$}
\label{ssec:o1}

Based on chemical reaction representing promoter assignment, increasing the value of $\omega_1$ decreases the number of active genes by 1, decreases the number of $xrn1$ complexes by 1, and increasing the number of $xrn1$ binding to promoters by 1. Decreasing the number of active genes results in reduced occurrence of transcription; decreasing the number of $xrn1$ complexes reduces promoter assignment in a feedback loop; increasing $xrn1$ binding to promoters increases $xrn1$-based transcription. Thus, the first effect (which does not directly involve $xrn1$) should decrease the number of proteins at steady state, while the last effect (which directly involves $xrn1$) should increase the number of proteins at steady state. Overall, as $\omega_1$ increases, the expected number of proteins at steady state decreases. This indicates that changing the number of active genes by a fixed amount has a greater effect on the number of proteins at steady state than does changing any of the $xrn1$ elements by the same fixed amount. In essence, although adding $xrn1$-based processes to the canonical model results in an effect, the basic processes present in both models still have the majority effect in determining steady state protein levels. Most of the change in protein levels happens between $\omega_1$ = 0.10 and 0.50.

\begin{table}
\begin{center}
	\begin{tabular}{|r| c c c c}
	\hline
	 & $1.25\rho_1$ & $1.50\rho_1$ & $1.75\rho_1$ & $2.00\rho_1$ \\
	\hline
	0.1 & 156.2 & 97.9 & 77.5 & 67.0 \\
	0.5 & 141.1 & 92.3 & 74.1 & 64.5 \\
	1.0 & 139.4 & 91.7 & 73.7 & 64.2 \\
	1.5 & 138.8 & 91.4 & 73.5 & 64.1 \\
	\end{tabular}

	\caption{\footnotesize{Expected number of proteins as given by deterministic solutions of the feedback model. The horizontal labels show varying values of $\omega_2$. The vertical labels show varying values of $\omega_1$. Much like the canonical model (Table~\ref{table:det-can}), deterministic solutions of the feedback model are independent of the value of $\gamma_1$. $\gamma_1$ = $0.01$, $0.05$, $0.14$, and $0.50$ resulted in the exact same values for expected number of proteins.}}
	
	\label{table:det-feed}
\end{center}
\end{table}

\begin{figure}
\begin{center}
	\includegraphics[width=\textwidth]{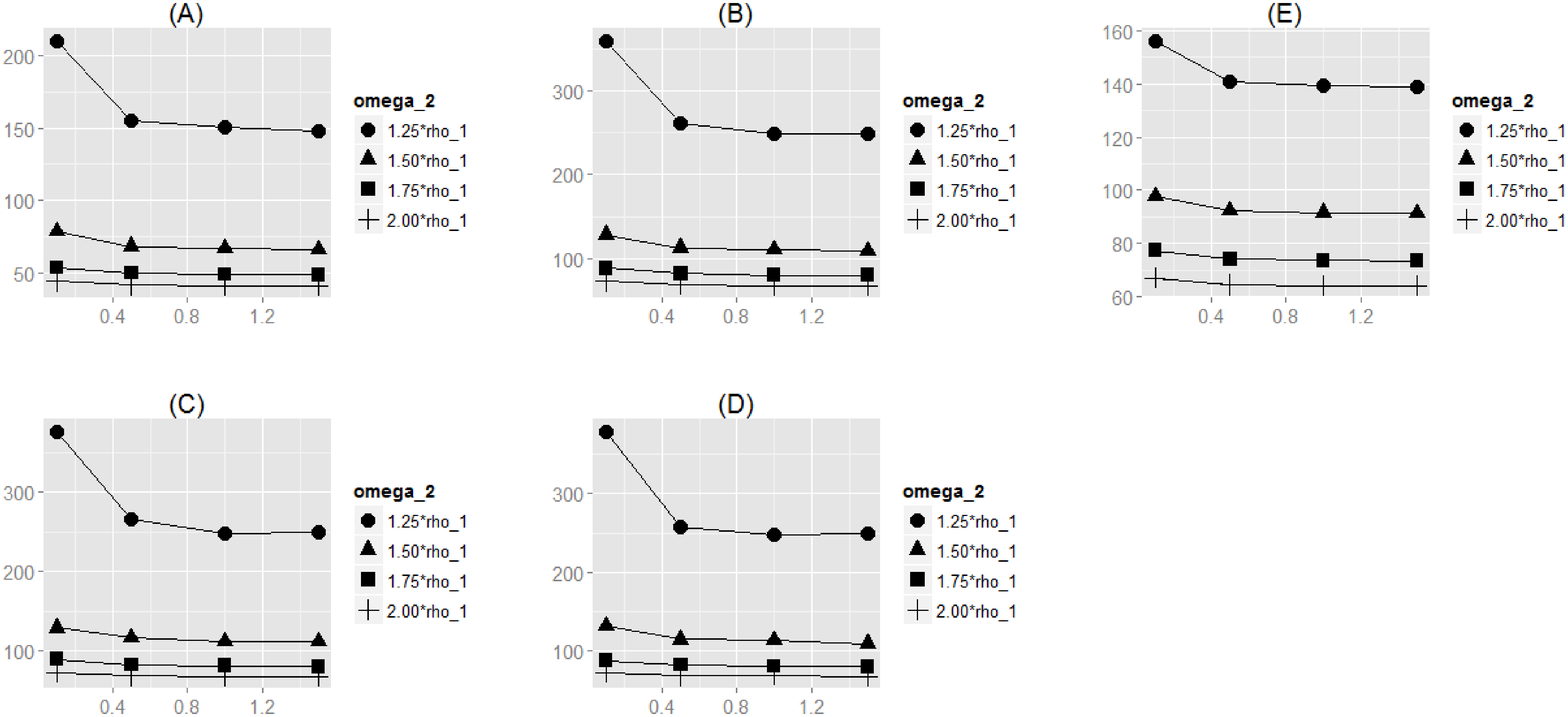}
	\caption{\footnotesize{Deterministic (E) and stochastic (A-D) solutions for number of proteins at steady state with the given parameter values: $\lambda_1$ =1; $\lambda_2$ = 15; $\rho_1$ = 0.0231. The x-axis on all plots is $\omega_1$; the y-axis on all plots is the mean number of proteins at steady state. (A) $\gamma_1$ = 0.01; (B) $\gamma_1$ = 0.05; (C) $\gamma_1$ = 0.14; (D) $\gamma_1$ = 0.50; (E) Deterministic solution is independent of the value of $\gamma_1$.}}
	\label{fig:vary-omega1}
\end{center}
\end{figure}

In addition to affecting the mean of steady state protein distribution, increasing $\omega_1$ also results in an overall decrease in the standard deviation of the steady state protein distribution. However, for larger values of $\gamma_1$ (e.g. $\gamma_1$ = 0.50, Table~\ref{table:stoch-feed0.50}) the standard deviation is either approximately equal to or larger than than the expected number of proteins at steady state. This indicates a high degree of variation in the steady state distribution. Therefore, even though exploring the effect of varying $\omega_1$ in the deterministic solution (Figure~\ref{fig:vary-omega1}E) shows the same trend as it does in the stochastic results, due to the relatively large values of the standard deviation, failing to account for variability in the data is a significant con of the deterministic solution.

\begin{table}
\begin{center}
	\begin{tabular}{|r| c c c c}
	\hline
	& $1.25\rho_1$ & $1.50\rho_1$ & $1.75\rho_1$ & $2.00\rho_1$ \\
	\hline
	0.1 & 210.3 $\pm$ 230.9 & 78.3 $\pm$ 40.8 & 53.3 $\pm$ 18.2 & 44.0 $\pm$ 12.8 \\
	0.5 & 155.2 $\pm$ 135.0 & 68.0 $\pm$ 27.9 & 49.5 $\pm$ 15.3 & 41.5 $\pm$ 11.6 \\
	1.0 & 150.2 $\pm$ 131.4 & 66.8 $\pm$ 26.7 & 48.6 $\pm$ 14.7 & 41.0 $\pm$ 11.5 \\
	1.5 & 147.8 $\pm$ 125.9 & 65.8 $\pm$ 25.3 & 48.5 $\pm$ 14.5 & 40.9 $\pm$ 11.2 \\ 
	\end{tabular}
	
	\caption{\footnotesize{Expected number of proteins $\pm$ standard deviation of the steady state protein distribution as given by stochastic simulations of the feedback model. The horizontal labels show varying values of $\omega_2$. The vertical labels show varying values of $\omega_1$. $\lambda_1/(\lambda_1+\lambda_2)=1/15$; $\rho_1=0.0231$; $\rho_2=\rho_1/10$; $\rho_3=\rho_1$; $\gamma_1=0.01$; $\gamma_2=\gamma_1/540$.}}

	\label{table:stoch-feed0.01}
\end{center}
\end{table}

\begin{table}
\begin{center}
	\begin{tabular}{|r| c c c c}
	\hline
	& $1.25\rho_1$ & $1.50\rho_1$ & $1.75\rho_1$ & $2.00\rho_1$ \\
	\hline
	0.1 & 359.9 $\pm$ 482.6 & 128.0 $\pm$ 87.9 & 88.3 $\pm$ 40.5 & 72.4 $\pm$ 27.3 \\
	0.5 & 261.6 $\pm$ 307.5 & 112.8 $\pm$ 62.0 & 81.9 $\pm$ 35.1 & 68.6 $\pm$ 24.6 \\
	1.0 & 249.5 $\pm$ 288.7 & 110.3 $\pm$ 63.0 & 79.9 $\pm$ 32.4 & 67.5 $\pm$ 23.9 \\
	1.5 & 248.8 $\pm$ 310.0 & 108.7 $\pm$ 58.3 & 80.1 $\pm$ 32.4 & 67.3 $\pm$ 23.7 \\ 
	\end{tabular}
	
	\caption{\footnotesize{Expected number of proteins $\pm$ standard deviation of the steady state protein distribution as given by stochastic simulations of the feedback model. The horizontal labels show varying values of $\omega_2$. The vertical labels show varying values of $\omega_1$. $\lambda_1/(\lambda_1+\lambda_2)=1/15$; $\rho_1=0.0231$; $\rho_2=\rho_1/10$; $\rho_3=\rho_1$; $\gamma_1=0.05$; $\gamma_2=\gamma_1/540$.}}

	\label{table:stoch-feed0.05}
\end{center}
\end{table}

\begin{table}
\begin{center}
	\begin{tabular}{|r| c c c c}
	\hline
	& $1.25\rho_1$ & $1.50\rho_1$ & $1.75\rho_1$ & $2.00\rho_1$ \\
	\hline
	0.1 & 376.6 $\pm$ 741.3 & 129.4 $\pm$ 137.1 & 89.3 $\pm$ 64.5 & 72.8 $\pm$ 44.1 \\
	0.5 & 265.7 $\pm$ 449.3 & 116.5 $\pm$ 107.2 & 82.2 $\pm$ 52.8 & 69.0 $\pm$ 38.4 \\
	1.0 & 247.7 $\pm$ 398.9 & 111.9 $\pm$ 98.8 & 81.3 $\pm$ 52.2 & 68.2 $\pm$ 36.9 \\
	1.5 & 249.0 $\pm$ 407.0 & 111.6 $\pm$ 98.9 & 80.4 $\pm$ 49.7 & 68.0 $\pm$ 38.9 \\
	\end{tabular}
	
	\caption{\footnotesize{Expected number of proteins $\pm$ standard deviation of the steady state protein distribution as given by stochastic simulations of the feedback model. The horizontal labels show varying values of $\omega_2$. The vertical labels show varying values of $\omega_1$. $\lambda_1/(\lambda_1+\lambda_2)=1/15$; $\rho_1=0.0231$; $\rho_2=\rho_1/10$; $\rho_3=\rho_1$; $\gamma_1=0.14$; $\gamma_2=\gamma_1/540$.}}

	\label{table:stoch-feed0.14}
\end{center}
\end{table}

\begin{table}
\begin{center}
	\begin{tabular}{|r| c c c c}
	\hline
	& $1.25\rho_1$ & $1.50\rho_1$ & $1.75\rho_1$ & $2.00\rho_1$ \\
	\hline
	0.1 & 379.1 $\pm$ 903.7 & 131.6 $\pm$ 209.7 & 87.7 $\pm$ 101.2 & 72.9 $\pm$ 72.5 \\
	0.5 & 257.1 $\pm$ 597.4 & 114.5 $\pm$ 164.4 & 81.9 $\pm$ 90.6 & 68.3 $\pm$ 65.8 \\
	1.0 & 248.5 $\pm$ 535.6 & 113.4 $\pm$ 161.9 & 80.7 $\pm$ 92.8 & 69.0 $\pm$ 67.2 \\
	1.5 & 249.0 $\pm$ 577.7 & 109.2 $\pm$ 156.8 & 79.6 $\pm$ 85.7 & 67.5 $\pm$ 65.8 \\ 
	\end{tabular}
	
	\caption{\footnotesize{Expected number of proteins $\pm$ standard deviation of the steady state protein distribution as given by stochastic simulations of the feedback model. The horizontal labels show varying values of $\omega_2$. The vertical labels show varying values of $\omega_1$. $\lambda_1/(\lambda_1+\lambda_2)=1/15$; $\rho_1=0.0231$; $\rho_2=\rho_1/10$; $\rho_3=\rho_1$; $\gamma_1=0.50$; $\gamma_2=\gamma_1/540$.}}

	\label{table:stoch-feed0.50}
\end{center}
\end{table}

\begin{figure}
\begin{center}
	\includegraphics[width=\textwidth]{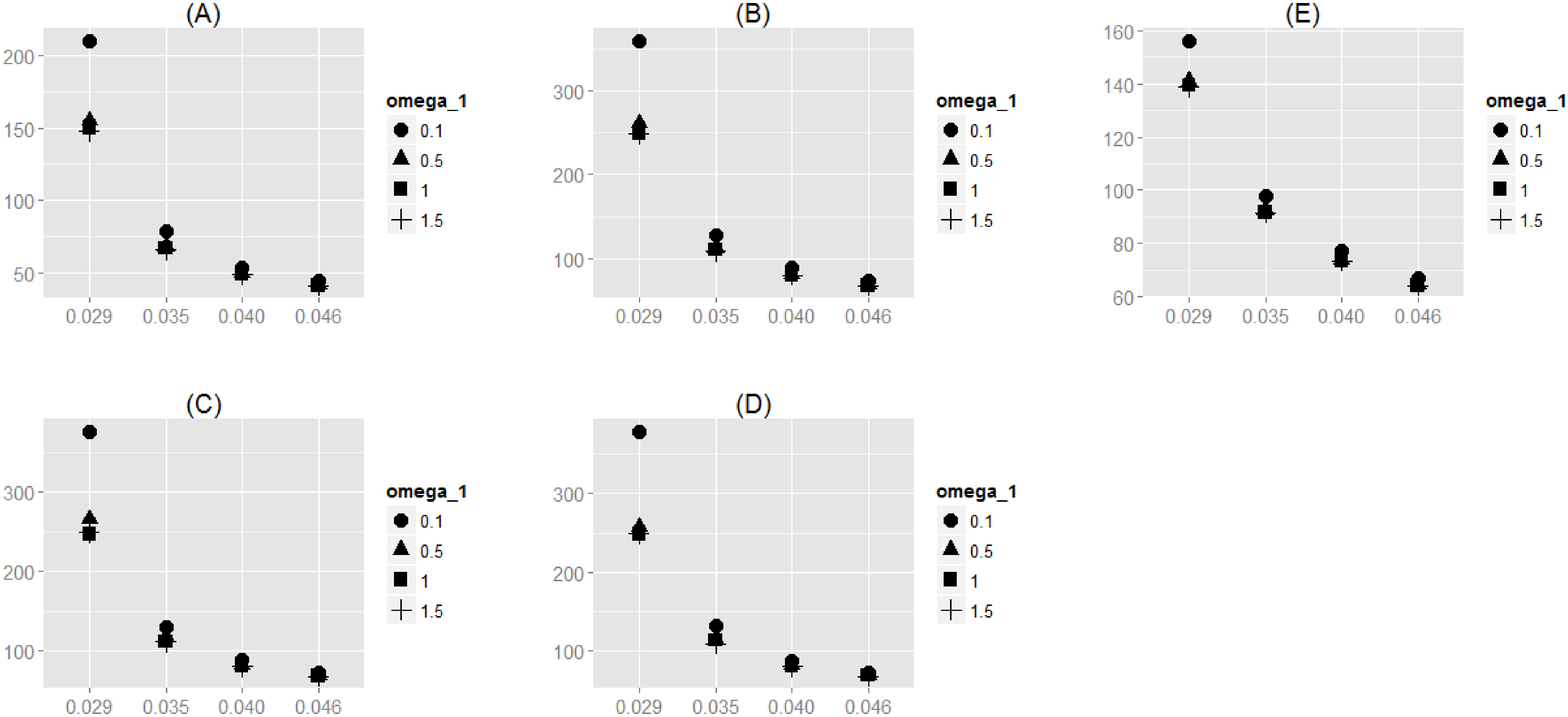}
	\caption{\footnotesize{Deterministic (E) and stochastic (A-D) solutions for number of proteins at steady state with the given parameter values: $\lambda_1$ =1; $\lambda_2$ = 15; $\rho_1$ = 0.0231. The x-axis on all plots is $\omega_2$; the y-axis on all plots is the mean number of proteins at steady state. (A) $\gamma_1$ = 0.01; (B) $\gamma_1$ = 0.05; (C) $\gamma_1$ = 0.14; (D) $\gamma_1$ = 0.50; (E) Deterministic solution is independent of the value of $\gamma_1$.}}
	\label{fig:vary-omega2}
\end{center}
\end{figure}

\subsection{Varying the rate of promoter reassignment: $\omega_2$}
\label{ssec:o2}

Varying $\omega_2$ has parallel qualitative results to varying $\omega_1$. Increasing the value of $\omega_2$ decreases the expected number of proteins at steady state, albeit at a different rate. Additionally, the largest change in the expected number of proteins occurs at lower values of $\omega_2$ (between $\omega_2$ = 1.25$\rho_1$ and 1.50$\rho_1$). Like $\omega_1$, increasing $\omega_2$ reduces the standard deviation of the steady state protein distribution (Tables~\ref{table:stoch-feed0.01}, \ref{table:stoch-feed0.05}, \ref{table:stoch-feed0.14}, \ref{table:stoch-feed0.50}), yet the standard deviation remains large enough relative to the expected value to merit the use of stochastic results over the deterministic solution. However, the deterministic solution does preserve the general trend observed in the stochastic results (Figure~\ref{fig:vary-omega2}). 

\section{Discussions}
\label{sec:discuss}

Stochastic chemical kinetics now has a renaissance due to the consequence of the emergence and development of systems biology. It looks to be one of the most important modeling tool to understand and describe the mechanism of gene expression. While it is one of the basic processes of life, we are far from having a detailed kinetic mechanism of the whole process composed of many subprocesses.
Generally a kinetic mechanism is said to be ``known'', if all elementary reactions and their rate constants are determined.

Genetic expression is modeled by lumped kinetic models. In a \emph{lumped} model, one step contains a sequence of more elementary reaction steps. The canonical model of genetic expression \cite{paulsson05} is technically a compartmental system, and its stochastic model can be completely solved. However, the incorporation of other steps of course implies changes in the kinetic properties of the system under investigation. More specifically, as it was stated recently ``protein distribution shape informs on molecular mechanism'' \cite{sherman-cohen14}. By following the same logic, we were interested in the qualitative (modality, skewness etc.) and quantitative features of the stationary distributions of different models. 

A comparative analysis of the canonical and a feedback model was given here. The construction of the feedback model has been motivated by the circular gene expression hypothesis \cite{haimovich13}, which assumes a mechanism of the interaction between the degradation and synthesis of mRNA. In the model we incorporated three lumped reactions, such as promoter assignment, promoter reassignment, and a second transcription step, which depend on the large protein $xrn1$. The collection and estimation of rate constants is not easy. The data used here is based on E. coli as a model organism, and the results could be different for eukaryotes and other organisms. There are initial encouraging results for obtaining more quantitative data~\cite{schwanh11,lee09,choudhary14} and there is a hope that it will be possible to give more reliable and consistent estimation of the rate constants. As concerns the analysis of the model, we restricted ourselves here for simulation studies and for the analysis of these results. Stationary distributions have been empirically constructed from the set of the individual realizations.

How to interpret the results? While our main goal was to see the whether there are characteristic differences between the canonical and the feedback models, remarkable effects of the some changes in the rate constants were also observed.
Most interestingly, the increase of translational rate in the canonical model destroys gamma distribution and leads to the emergence of some kinds of multimodality. It is important to note that the corresponding deterministic model leads to uni-stationarity (and not multi-stationarity). As the realizations show the transient behavior, a system is generally in one of the two possible ``high'' and ``low'' states with rapid jumps between them. In the canonical model we don't see multimodality, but exponential distribution was fitted well.
Increased transcription rate implies more expressed right-skewness, in both model, while increased values of the promoter reassignment rate  result in a decreased expected value of the protein distribution.  The systematic exploration of the three-dimensional parameter space of the rates of the additional reactions of the feedback model should be the next step.

In summary, our studies support the view that qualitative and quantitative changes in the shape and in the numerical characteristics of the stationary distributions of the stochastic models occur due to the consequence of altered reaction network and rate constants. Combined experimental and theoretical studies could help to uncover the details of the kinetic mechanism of the circular gene hypothesis.

\section*{Acknowledgements}
\label{sec:ack}

Thanks for Mordechai (Motti) Choder for motivation and initial correspondence. PE thanks to the Henry Luce Foundation to let him to serve as a Henry R Luce Professor.

\pagebreak

\bibliography{cirr}

\end{document}